\documentclass[reprint,superscriptaddress,amsmath,amssymb,aps,notitlepage,floatfix,longbibliography]{revtex4-1}
\usepackage{graphicx}
\usepackage{color}
\usepackage{comment}

\usepackage{dcolumn}
\graphicspath{{./Figures_ASHE}}
\usepackage{upgreek}
\usepackage{gensymb}
\usepackage{float}
\usepackage{bm}
\usepackage{xcolor}
\usepackage{physics}

\usepackage{multirow}
\usepackage{array}

\usepackage{ulem}

\usepackage{xr}
\makeatletter
\newcommand*{\addFileDependency}[1]{
	\typeout{(#1)}
	\@addtofilelist{#1}
	\IfFileExists{#1}{}{\typeout{No file #1.}}
}
\makeatother

\begin{document}

\title{Electromagnetic evanescent field associated with surface acoustic wave: \\ 
	Response of metallic thin films} 

\author{Takuya Kawada}
\email[]{takuyakawada@g.ecc.u-tokyo.ac.jp}
\affiliation{Department of Physics, The University of Tokyo, Bunkyo, Tokyo 113-0033, Japan}
\affiliation{Department of Basic Science, The University of Tokyo, Meguro, Tokyo 153-8902, Japan}

\author{Kei Yamamoto}
\email[]{yamamoto.kei@jaea.go.jp}
\affiliation{Advanced Science Research Center, Japan Atomic Energy Agency, Tokai, Ibaraki 319-1195, Japan}
\affiliation{RIKEN Center for Emergent Matter Science, Wako, Saitama 351-0198, Japan}

\author{Masashi Kawaguchi}
\affiliation{Department of Physics, The University of Tokyo, Bunkyo, Tokyo 113-0033, Japan}

\author{Hiroki Matsumoto}
\affiliation{Department of Physics, The University of Tokyo, Bunkyo, Tokyo 113-0033, Japan}
\affiliation{Institute for Chemical Research, Kyoto University, Uji, Kyoto 611-0011, Japan}

\author{Ryusuke Hisatomi}
\affiliation{Institute for Chemical Research, Kyoto University, Uji, Kyoto 611-0011, Japan}
\affiliation{Center for Spintronics Research Network (CSRN), Kyoto University, Uji, Kyoto 611-0011, Japan}

\author{Hiroshi Kohno}
\affiliation{Department of Physics, Nagoya University, Chikusa, Nagoya 464-8602, Japan}

\author{Sadamichi Maekawa}
\affiliation{Advanced Science Research Center, Japan Atomic Energy Agency, Tokai, Ibaraki 319-1195, Japan}
\affiliation{RIKEN Center for Emergent Matter Science, Wako, Saitama 351-0198, Japan}

\author{Masamitsu Hayashi}
\affiliation{Department of Physics, The University of Tokyo, Bunkyo, Tokyo 113-0033, Japan}
\affiliation{Trans-Scale Quantum Science Institute (TSQS), The University of Tokyo, Bunkyo, Tokyo 113-0033, Japan}

\newif\iffigure
\figurefalse
\figuretrue

\date{\today}

\begin{abstract}
	Surface acoustic waves (SAWs), when excited using a piezoelectric material, generate not only strain but also an electric field at the surface. Conventional analyses of the electric field associated with SAWs rest on the quasi-electrostatic approximation, which neglects the accompanying magnetic field and may thus fail to fully capture some physical properties of SAWs. In this study, we investigate the electric and magnetic fields induced by SAWs without introducing the quasi-electrostatic approximation. The system of interest is a piezoelectric substrate covered with a thin conductive layer. The plane wave solution, with phase velocity equal to the speed of sound, takes the form of an evanescent field with a non-negligible transverse component. The transverse electric field penetrates the conductive layer, inducing time- and space-dependent current that flows uniformly across the film thickness. The current modifies and is screened by the evanescent magnetic field, which can be comparable in magnitude to the Barnett field associated with the SAW-induced rotational motion of the surface in highly conductive films. The transverse electric field reduces to a harmonic one in the quasi-electrostatic limit, elucidating the mechanism by which the latter evades electrostatic screening. 
    This work provides a comprehensive framework for evaluating SAW-induced evanescent fields, offering a clear physical interpretation of the unscreened electric field and the associated magnetic field.
\end{abstract}

\maketitle

\section{Introduction}
Surface acoustic waves (SAWs) are vibrational modes localized at the surfaces of solids.
SAWs are typically excited on a piezoelectric material and thus accompanied by a time- and space-dependent electric field~\cite{white1965apl}.
In condensed matter physics, SAWs have been used as effective tools for non-invasive manipulation and detection of the electronic states.
Materials of interest are placed on the piezoelectric substrate where the SAW propagates. 
The SAW-induced ac electric field mobilizes charge carriers in (semi-)conducting materials, which in turn, alter the SAW's amplitude and velocity \textit{via} a back reaction effect~\cite{collins1968apl,ingebrigtsen1970jap,ricco1985,wixforth1986}. 
Measuring the amplitude and velocity of SAWs thus provides a means to probe the electronic states of materials in the microwave frequency range,
offering a robust technique for tracking low-frequency conductivity
~\cite{adler1981apl,fritzche1984prb,paalanen1992prb,karl2000,muller2005jap,wu2024prl} and diagnosing quantum Hall states~\cite{efros1990prl,falko1993,rotter1998apl,fang2023}. 

Recently, SAWs have attracted growing interest in the field of spintronics for controlling the electron spin.
That is, the mechanical motion of lattice can interact with the spin of free carriers \textit{via} spin-vorticity coupling~\cite{matsuo2013prb,kobayashi2017prl,mingxian2023prb} or with the magnetization of magnetic materials through magneto-elastic coupling~\cite{weiler2011prl,dreher2012prb,thevenard2014prb,sasaki2017prb}.
In non-magnetic materials, spin-vorticity coupling is thought to be a unique mechanism of SAW-spin current conversion that is independent of spin-orbit interaction.
On the other hand, a different type of spin current in non-magnetic metals has been reported, which requires strong spin-orbit interaction~\cite{kawada2021sciadv}.
Although this spin current had been suggested to be of a mechanical origin, later studies indicated otherwise \textit{i.e.}, it is of an electrical origin~\cite{kawada2024exparxiv}.
Recent study has reported yet another coupling \textit{via} magnetic field generated by the electric field inductively~\cite{kline2024prap}.
It is thus important to clarify the electromagnetic character of SAWs to distinguish mechanical contribution from the electrical one to the physical effects in both magnetic and non-magnetic materials.


Conventional analysis of the ac electric field $\bm{E}$ associated with SAWs in piezoelectric materials introduces a scalar potential $\phi$ and sets $\bm{E}=-\grad{\phi}$~\cite{tiersten1963jasa,tseng1967jap,campbell1968ieee,ingebrigtsen1969jap}. When applied to time-dependent problems beyond the \textit{electrostatics}, which is exact only for strictly time-independent problems, this ansatz is sometimes called \textit{quasi-electrostatic approximation}~\cite{jackson1998}. In 1970, Ingebrigtsen calculated the ac electric field that emanates from a propagating SAW in a piezoelectric substrate and permeates into a neighboring conducting film~\cite{ingebrigtsen1970jap}. The study, based on the quasi-electrostatic approximation, showed that the electric field in the conductor comprises two components, each decaying exponentially away from the interface. 
One component is confined near the interface within the Thomas-Fermi length $\lambda _{\rm TF} \sim$~0.1~nm for a metal, and the other component remains nearly uniform across the film thickness, as long as the film is thinner than the SAW  wavelength, typically of the order of 1--10~$\upmu$m. 
This result implies that electric field is not fully screened by free charges, even in the treatment based on electrostatics. It leaves the physics governing the unscreened component unclear since its decay length is independent of any material property.

In the quasi-electrostatic approximation, Faraday's law forces the magnetic field $\bm{H}$ to be constant in time, assuming the system is non-magnetic \textit{i.e.}, $\bm{B}=\mu _0 \bm{H}$ where $\mu _0$ is the permeability of vacuum.  It is often thought to be quantitatively accurate if the retardation effect can be ignored \textit{i.e.}, if the characteristic time scale of the problem is much longer than the time it takes for the light to travel past the system. In the context of SAW-related applications, the condition typically translates to the smallness of the SAW phase velocity $v$ compared to the speed of light $c$, which is obviously very well satisfied, usually by five orders of magnitude. Based on this reasoning, $\bm{H}$ is practically removed from the discussions, and specifically the Amp\'{e}re-Maxwell law ends up being entirely disregarded. 
Indeed, this is the argument made in Ref.~\cite{ingebrigtsen1970jap} to justify the approximation used.
It is not immediately clear, however, whether $v/c\ll1 $ is \textit{sufficient} for neglecting $\bm{H}$, especially in materials characterized by additional parameters carrying space-time dimensions. 
In highly conductive metals, this question is related to estimating the amplitude of electric currents, where those associated with $\phi $ and $\bm{H}$ are screened by distinct physical mechanisms.

In this work, we study electric and magnetic fields accompanying SAWs in a piezoelectric substrate covered by a conducting film without invoking the quasi-electrostatic approximation. The plane wave solution takes the form of an evanescent field that propagates at the sound velocity and permeates into the film with two characteristic decay constants, $\lambda _{\rm TF}$ and a generalized skin depth corresponding to the longitudinal and transverse screening mechanisms respectively (Secs.~\ref{sec:bulk} and \ref{sec:full_film}). 
We show that the transverse fields are crucial for satisfying the boundary conditions, and responsible for the dominant component of the electric current that flows in the conducting layer uniformly even when $v/c\ll1$. Despite the counterintuitive relevance of the transverse components, the solution does reduce to that obtained in the quasi-electrostatic approximation upon taking an appropriate limit. This allows us to identify the conditions for which the ansatz $\bm{E}=-\nabla \phi $ is valid and to unambiguously designate a part of $\phi $, often referred to as harmonic potential, to be transverse (Sec.~\ref{sec:electrostatics}). With the quasi-electrostatic approximation validated under the relevant conditions, we present quantitative estimation of the transverse magnetic field and the associated electric current generated by SAWs in the conducting film (Sec.~\ref{sec:estimates}). This work thus bridges the gap between the general framework for the electromagnetic response of free carriers in conductive films and the simplified treatment under the quasi-electrostatic approximation commonly employed in SAW studies, offering a clear physical interpretation of the unscreened electric field and establishing a quantitative evaluation of the SAW-induced electromagnetic field.

\section{Setup}
\begin{figure}[b]
	\centering
	\begin{minipage}{1.0\hsize}
		\centering
		\includegraphics[scale=0.06]{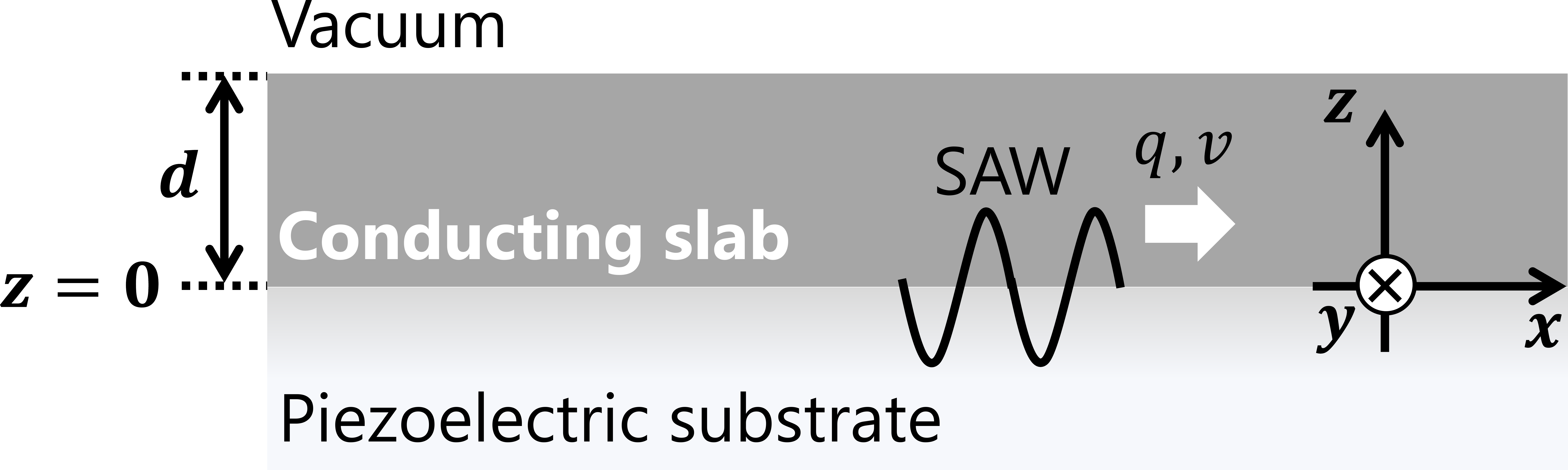}
	\end{minipage}
	\caption{
		Schematic illustration of the system, composed of a piezoelectric substrate ($z<0$), a conducting layer  ($0<z<d$), and the vacuum ($z>d$).
		A piezoelectrically excited SAW, with wavenumber $q$ and phase velocity $v$ along $x$, travels on the surface of the substrate.
	}
	\label{fig:calc_Fig1_rev}
\end{figure}

We study plane wave propagation along the interface between a piezoelectric substrate and a (nonmagnetic) conducting layer of thickness $d$ facing vacuum. 
Let the $xyz$ coordinate system be oriented such that the wave propagates along $x$ and the substrate occupies $z<0$ as shown in Fig.~\ref{fig:calc_Fig1_rev}.
We would like to know the electromagnetic waves described by the macroscopic Maxwell's equations, 
\begin{eqnarray}
	\partial _t \bm{B} &=& -\nabla \times \bm{E} , \label{eq:Maxwell1} \\
	0 &=& \nabla \cdot \bm{B} , \label{eq:Maxwell2} \\
	\partial _t \bm{D} &=& \nabla \times \bm{H} -\bm{J} , \label{eq:Maxwell3}  \\
	0 &=& \nabla \cdot \bm{D} - \rho , \label{eq:Maxwell4}
\end{eqnarray}
where $\rho $ denotes the free charge density. Different material media are distinguished by the constitutive relations between $\bm{E} ,\bm{H}$ and $\bm{D},\bm{B}$ as well as the assumptions on the free charges. We  model each region as follows: 
\begin{description}
	\item[ Vacuum ($z>d$)] 
	\begin{equation}
		\bm{D} = \varepsilon _0 \bm{E} , \quad \bm{B} = \mu _0 \bm{H} , \quad \rho = 0 , \quad \bm{J} = \bm{0} ,
	\end{equation}
	where $ \varepsilon _0$ is the permittivity of vacuum.
	\item[ Conducting layer ($0<z<d$)] 
	\begin{equation}
		\begin{gathered}
			\bm{D}=\varepsilon _r \varepsilon _0 \bm{E} , \quad \bm{B} = \mu _0 \bm{H} , \\
			\partial _t \rho +\nabla \cdot \bm{J} = 0 , \quad \bm{J} = \sigma _c \bm{E} - D_e \nabla \rho ,
		\end{gathered} \label{eq:isotropic_conductor} 
	\end{equation}
	where $\varepsilon _r , \sigma _c , D_e$ are the relative permittivity, conductivity, and the diffusion constant, respectively.
	\item[ Piezoelectric substrate ($z<0$)]
	\begin{equation}
		\bm{D} = \hat{\varepsilon } \bm{E} + \bm{P} , \quad \bm{B} = \mu _0 \bm{H} , \quad 
		\rho = 0 , \quad \bm{J} =\bm{0} , \label{eq:piezo_relation}
	\end{equation}
	where $\hat{\varepsilon }$ is the dielectric tensor, and $\bm{P}= \hat{e} \hat{\epsilon } $ is the electric polarization associated with the linear strain tensor $\hat{\epsilon } $  through the piezoelectric tensor $\hat{e}$. 
	The displacement vector $\bm{u}$, related to the strain by $\epsilon _{ij} = (\partial _j u_i + \partial _i u_j )/2$, obeys the equation of motion, 
	\begin{equation}
		m_\mathrm{p} \partial ^2_t u_i = c_{ijkl} \partial _j \partial _l u_k -e_{kij}  \partial _j E_k ,  \label{eq:elastic}
	\end{equation}
	where $m_\mathrm{p}$ is the mass density of the piezoelectric substrate and $c_{ijkl}$ is the elastic stiffness tensor. 
	Repeated Latin indices are understood to be summed over. 
\end{description}
Since all the equations involved are linear, the solutions can be sought in the form $f\left( z\right) e^{iq\left( x-vt \right)} $ where $q$ and $v$ are the in-plane wavenumber and phase velocity, respectively. To avoid notational mess, we hereafter understand all the dependent variables are of this form and suppress the appearance of $e^{i q(x - vt)}$.
Under our assumption that $\bm{B}=\mu _0 \bm{H}$ holds everywhere, we can eliminate $\bm{B}$ and $\bm{H}$ \textit{via} Eqs.~(\ref{eq:Maxwell1}) and (\ref{eq:Maxwell2}), 
\begin{equation}
	i qv \begin{pmatrix}
		B_x \\
		B_y \\
		B_z \\
	\end{pmatrix} = \begin{pmatrix}
		-\partial _z E_y \\
		\partial _z E_x -i qE_z \\
		iqE_y \\
	\end{pmatrix} , \quad \partial _z B_z = -iqB_x .\label{eq:magnetic} 
\end{equation}
Note that Eq.~(\ref{eq:Maxwell2}) is redundant for $\omega \equiv qv \neq 0$.
From Eqs.~(\ref{eq:Maxwell1}) and (\ref{eq:Maxwell3}), one obtains
\begin{equation}
	\begin{gathered}
		q^2 v^2  \mu _0 \begin{pmatrix}
			D_x \\
			D_y \\
			D_z \\
		\end{pmatrix} = \begin{pmatrix}
			-\partial^2_z E_x + iq \partial _z E_z  \\
			\qty(q^2 -\partial _z^2) E_y \\
			iq \partial _z E_x + q^2 E_z \\
		\end{pmatrix} -i qv \mu _0 \begin{pmatrix}
			J_x \\
			J_y \\
			J_z \\
		\end{pmatrix} . \\
		\label{eq:TM_TE}
	\end{gathered}
\end{equation}
The constraint Eq.~(\ref{eq:Maxwell4}) becomes redundant for $\omega \rho \neq 0$ because $\rho $ and $\bm{J}$ satisfy the conservation law.
The equations are supplemented by boundary conditions appropriate for different material interfaces. 
For book-keeping, we categorize them into three classes: (i) those imposing conditions on electronic or mechanical degrees of freedom, (ii) those imposed at spatial infinities $z \rightarrow \pm \infty $, and (iii) the continuity of appropriate components of $\bm{E},\bm{H},\bm{D},\bm{B}$ at interfaces, referred to as electromagnetic boundary conditions hereafter. They are all independent of each other, at least for the cases of our interest. In the following section, we first solve the equations in each region and impose the boundary conditions of types (i) and (ii) which can be discussed independently of the other regions. Then in Sec.~\ref{sec:full_film}, we glue the solutions together at the interfaces by using the remaining electromagnetic boundary conditions (iii).

\section{Solution in each region}\label{sec:bulk}
In this section, we solve Eq.~(\ref{eq:TM_TE}) in each region separately.
The obtained solutions will be conjoined by the electromagnetic boundary conditions in the next section.

\subsection{Vacuum half-space}
First we consider the vacuum half-space ($z>d$). 
We demonstrate that the evanescent field associated with SAW is a universal feature of vacuum electromagnetic waves forced to propagate slower than the speed of light $c$. 
Noting $\mu _0 \varepsilon _0 =1/c^2 $, Eq.~(\ref{eq:TM_TE}) yields
\begin{eqnarray}
	\frac{ q^2 v^2}{c^2}\begin{pmatrix}
		E_x \\
		E_y \\
		E_z \\
	\end{pmatrix} &=& \begin{pmatrix}
		-\partial _z^2 E_x +iq\partial _z E_z \\
		\left( q^2 -\partial _z^2 \right) E_y \\
		iq\partial _z E_x +q^2 E_z \\
	\end{pmatrix}  .  \label{eq:transverse_vacuum}
\end{eqnarray}
Imposing the vanishing boundary condition at $z\rightarrow \infty $, one obtains from Eq.~(\ref{eq:transverse_vacuum}),  
\begin{eqnarray}
	\begin{pmatrix}
		E_x \\
		E_z \\
	\end{pmatrix}
	&=& C^{\rm TM}
	\begin{pmatrix}
		1 \\
		i\gamma \\
	\end{pmatrix}
	e^{-q\left( z-d\right) /\gamma }, \label{eq:TM_mode}\\
	E_y
	&=&
	C^{\rm TE} e^{-q\left( z-d\right) /\gamma }, \label{eq:TE_mode}
\end{eqnarray}
where $\gamma =1/\sqrt{1-v^2 /c^2}$ is the Lorentz factor, and we assume $q > 0$ in the following.
The integration constants, $C^{\rm TE}$ and $C^{\rm TM}$, will be determined by the boundary conditions at  $z=d$.
Equations~(\ref{eq:TM_mode}) and (\ref{eq:TE_mode}) represent the transverse magnetic (TM) and the transverse electric (TE) modes, respectively.

\begin{figure}[b]
	\centering
	\begin{minipage}{1.0\hsize}
		\centering
		\includegraphics[scale=0.08]{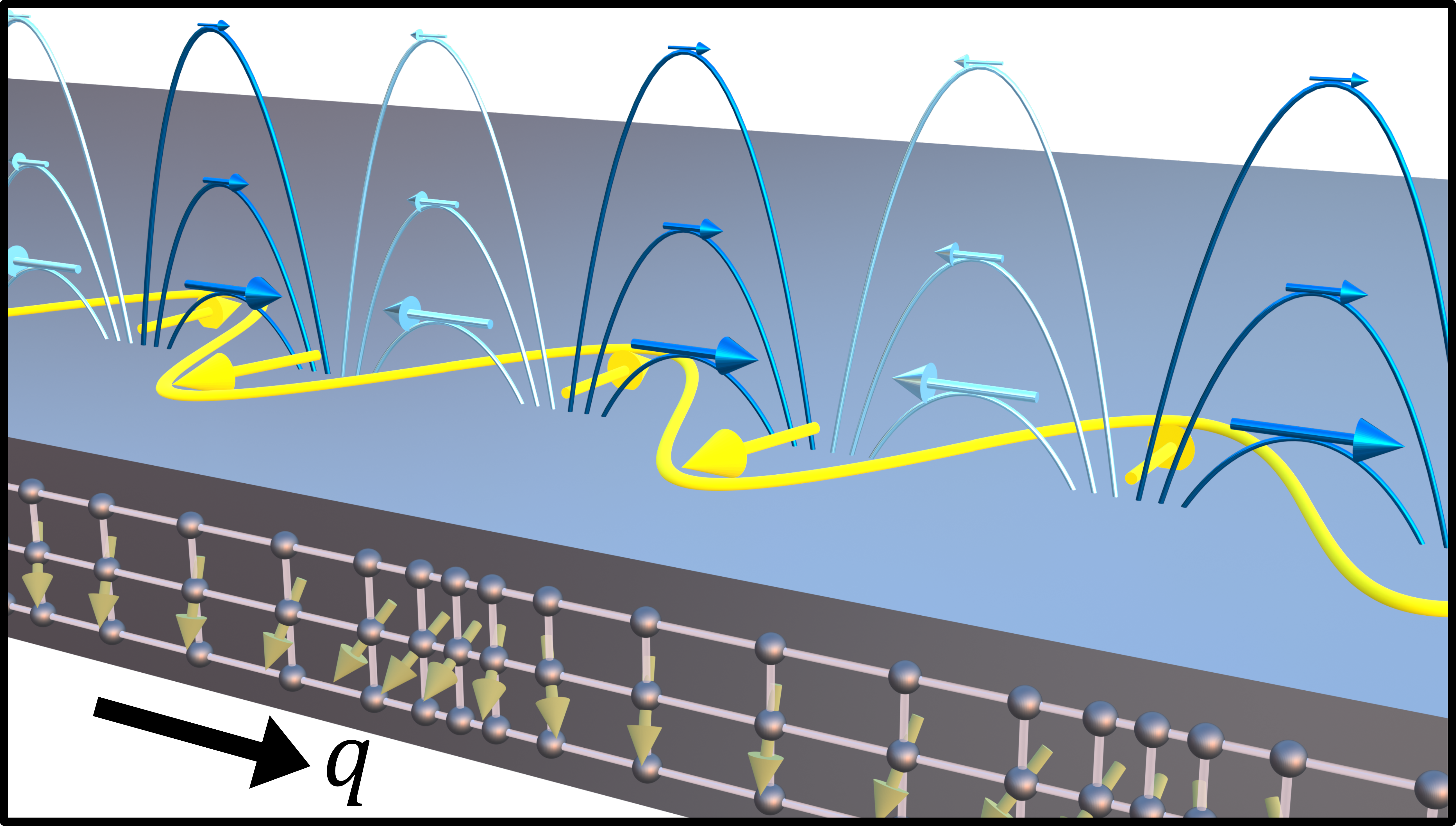}
	\end{minipage}
	\caption{
		Schematic illustration of the SAW-induced TM mode electromagnetic evanescent field. The purple and yellow curves with arrows represent the electric force and magnetic field, respectively. The length of the arrows indicate the strength of the coresponding fields. The green arrows show the local spontaneous polarization in the piezoelectric (e.g., LiNbO$_3$) substrate. The dense gray spheres represent lattice points of the substrate.
	}
	\label{fig:calc_Fig2_rev}
\end{figure}
The above solution has been derived independently of the boundary conditions at $z=d$. 
This shows that any electromagnetic waves in vacuum that decay at infinity $(z\rightarrow \infty)$ automatically take the form of the 
evanescent wave~\cite{bliokh2014}, in the sense that its Poynting vector is tangent to the interface, as was pointed out for the special case of the Bleustein-Glyaev-type SAW~\cite{bleustein1968apl,gulyaev1969jetp} in highly symmetric piezoelectric materials~\cite{li1996jap}.
As an example, the TM mode Eq.~(\ref{eq:TM_mode}) is schematically illustrated in Fig.~\ref{fig:calc_Fig2_rev}.
The electric and magnetic field lines display the presence of an electromagnetic field associated with the SAW.
Note that going beyond the quasi-electrostatic approximation predicts a nonzero magnetic field, hence the transverse component of the electric field, associated with the SAW.
Inserting Eq.~(\ref{eq:TM_mode}) into Eq.~(\ref{eq:magnetic}) gives $|B_y| = (\gamma v/c^2)|E_x|$. Using this relation, we obtain $|B_y|\sim 40$ nT if we assume $v=4000$ m/s and $|E_x|=$ 1 V/$\upmu$m, which corresponds to SAW strain $|\epsilon_{xx}|=2\times 10^{-4}$ according to numerical calculation of the SAW propagating in Y$+128^\circ$-cut LiNbO$_3$ substrate along the crystal $X$ axis~\cite{tiersten1963jasa,tseng1967jap,campbell1968ieee,ingebrigtsen1969jap,kawada2025jap}.
Although the magnetic field component is significantly smaller compared to the free electromagnetic field, this result can potentially be relevant because it directly couples to electron spin and magnetization~\cite{zhengzhi2023,kline2024prap}. 
We will discuss this in detail in Sec.~\ref{sec:estimates}.



\subsection{Conducting layer \label{sec:3b}} 

Next we consider the solution inside a conducting layer. 
We assume that the conductor is isotropic, and introduce the reduced speed of light $c_m = (\mu _0 \varepsilon _0 \varepsilon _r)^{-1/2}$, the skin depth $\delta_m = \sqrt{2/(qv\mu_0\sigma_c)}$, and the Thomas-Fermi length $\lambda_{\mathrm{TF}} = \sqrt{D_e\varepsilon_0 \varepsilon_r/\sigma_c}$. From Eq.~(\ref{eq:TM_TE}), one then obtains 
\begin{widetext}
	\begin{equation}
		\left( \frac{q^2 v^2}{c_m^2} +\frac{2i}{\delta_m^2} \right) \begin{pmatrix}
			E_x \\
			E_y \\
			E_z \\
		\end{pmatrix} \\
		=
		\begin{pmatrix}
			-\partial _z^2 E_x + iq\partial _z E_z \\
			\left( q^2 -\partial _z^2 \right) E_y \\
			iq\partial _z E_x + q^2 E_z  \\
		\end{pmatrix}
		+\frac{2i \lambda_{\mathrm{TF}}^2}{\varepsilon_0 \varepsilon_r \delta_m^2}
		\begin{pmatrix}
			iq\rho  \\
			0 \\
			\partial _z \rho  \\
		\end{pmatrix} , 
	\end{equation}
\end{widetext}
with
\begin{equation}
	\rho=   \varepsilon _0 \varepsilon _r \left( iqE_x + \partial _z E_z \right) . \label{eq:charge_density_metal}
\end{equation}
The general solutions are written in the form, 
\begin{equation}
	\bm{E} = \bm{E}^{\rm TM} + \bm{E}^{\rm TE} + \bm{E}^{\rm L} ,   
\end{equation}
where $\nabla \cdot \bm{E}^{\rm TM} = \nabla \cdot \bm{E}^{\rm TE}=0$,  
$\nabla \times \bm{E}^{\rm L}=\bm{0}$, and 
\begin{equation}
	\begin{aligned}
		\begin{pmatrix}
			E_x^{\rm TM} \\
			E_z^{\rm TM}  \\
		\end{pmatrix} =&\  C_+^{\rm TM}
		\begin{pmatrix}
			1  \\
			-iq/\alpha \\
		\end{pmatrix} e^{\alpha \left( z-d\right) }
		+ C_-^{\rm TM}
		\begin{pmatrix}
			1 \\
			iq/\alpha\\
		\end{pmatrix} e^{-\alpha z} , \\
		E_y^{\rm TE} =&\  C^{\rm TE}_+ e^{\alpha \left( z-d\right) } + C^{\rm TE}_- e^{-\alpha z} , \\
		\begin{pmatrix}
			E_x^{\rm L} \\
			E_z^{\rm L} \\
		\end{pmatrix}=&\ C_+^{\rm L}
		\begin{pmatrix}
			iq/Q \\
			1  \\
		\end{pmatrix} e^{Q \left( z-d\right) }
		+ C_-^{\rm L}
		\begin{pmatrix}
			-iq/Q \\
			1  \\
		\end{pmatrix} e^{-Qz}  ,
		\label{eq:E_metal_solved}
	\end{aligned}
\end{equation}
with
\begin{align}	
	\alpha &= q\sqrt{1-\frac{v^2}{c_m^2} -\frac{2i}{\left( q\delta _m \right) ^2}}, \label{eq:alpha}\\
	Q &= \frac{1}{\lambda _{\rm TF}} \sqrt{1 + \left( q\lambda _{\rm TF} \right) ^2 -i\eta  }, \label{eq:Q}\\
	\eta &= \frac{qv\varepsilon _0 \varepsilon _r}{ \sigma _c}  =   \frac{v^2}{2c_m^2} \left( q\delta _m \right) ^2. \label{eq:eta}
\end{align}
Note that the other components vanish: $ E^{\rm TM}_y = E^{\rm L}_y = E^{\rm TE}_x = E^{\rm TE}_z = 0$. 
Thus, the electric field consists of TM, TE, and longitudinal components with two integration constants each. 
In Eq.~(\ref{eq:eta}), $\eta$ represents the ratio of displacement current ($\varepsilon_r \varepsilon_0 \partial_t E$) to drift current ($\sigma_c E$), which is several orders of magnitude smaller than unity in metals for microwave frequency field.
For later use, we write down the electric current density, charge density, and magnetic flux density in the following:
\begin{align}
	\begin{pmatrix}
		J_x \\
		J_z \\
	\end{pmatrix} &= \sigma _c \begin{pmatrix}
		E^{\rm TM}_x \\
		E^{\rm TM}_z \\
	\end{pmatrix} + i \eta \sigma _c \begin{pmatrix}
		E^{\rm L}_x \\
		E^{\rm L}_z \\
	\end{pmatrix}, 	\  J_y = \sigma_c E_y^{\rm TE}\label{eq:Jf_result} , \\
	\rho & = -i\varepsilon _0 \varepsilon _r \frac{Q^2}{q}\left( 1 - \frac{q^2}{Q^2} \right) E_x^{\mathrm{L}}, \label{eq:rhof_result}
\end{align}
\begin{align}
		\begin{pmatrix}
			B_x \\
			B_z \\
		\end{pmatrix} &= \frac{1}{v} \left\{ 
		C^{\rm TE}_+
		\begin{pmatrix}
			i\alpha/q \\
			1 \\
		\end{pmatrix}e^{\alpha \left( z-d\right) }
		+ C^{\rm TE}_-
		\begin{pmatrix}
			-i\alpha/q \\
			1 \\
		\end{pmatrix}e^{-\alpha z} \right\} , \nonumber \\
    	B_y &= -\frac{1}{v} \left\{ 1 - \left( \frac{\alpha}{q} \right) ^2 \right\} E_z^{\rm TM} .
	\label{eq:B_result}
\end{align}

Without relying on the quasi-electrostatic approximation, we have obtained in Eq.~(\ref{eq:alpha}) the precise decay length $\alpha^{-1} $ of the transverse components, which originates from the Amp\'{e}re-Maxwell law [Eq.~(\ref{eq:Maxwell3})] and can be considered a generalized skin depth. 
Equation~(\ref{eq:alpha}) reproduces the previous result $\alpha=q$~\cite{ingebrigtsen1970jap} 
if we neglect the skin effect ($\delta _m \to \infty$) and take the limit $v/c\rightarrow 0$,  
whereas it reduces to the ordinary skin effect if we set $v = c_m$. 
Since the SAW velocity $v$ is about five orders of magnitude smaller than the speed of light, $v/c_m \ll 1$ is always satisfied.
Thus the difference between the updated $\alpha$ [Eq.~(\ref{eq:alpha})] and the quasi-electrostatic one~\cite{ingebrigtsen1970jap} comes from the imaginary term under the square root.
The derived expression suggests that the contribution of the skin effect cannot be neglected when $\delta_m$ is comparable to $q^{-1}$, in which case it predicts a shorter decay length than the conventional model.
Such a situation occurs in materials with sufficiently large conductivity and thus the result obtained here can be important especially for conducting layers. 

Let us proceed to discuss the boundary conditions that can be imposed independently of the other regions. 
Equation~(\ref{eq:E_metal_solved}) has six integration constants corresponding to three dynamical degrees of freedom; two for photons and another for conduction electrons. 
Two of the constants can be eliminated by imposing $J_z =0$ at the boundaries, enabling one to express $C_{\pm }^{\rm L}$ in terms of $C_{\pm }^{\rm TM}$;
\begin{equation}
	\begin{aligned}
		\begin{pmatrix}
			C_+^{\rm L} \\
			C_-^{\rm L} \\
		\end{pmatrix}
		=&\  \frac{qe^{Qd}}{2\alpha\eta  \sinh Qd}\\
		&\times\begin{pmatrix}
			1 - e^{-\left( Q+\alpha \right) d} & -e^{-\alpha d} + e^{-Qd} \\
			e^{-\alpha d} - e^{-Qd} & -1 + e^{-\left( Q+\alpha \right) d} \\
		\end{pmatrix} \begin{pmatrix}
			C_+^{\rm TM} \\
			C_-^{\rm TM} \\
		\end{pmatrix} . \label{eq:current_solved} 
	\end{aligned}
\end{equation}
Thus the general solutions in the conducting layer have four constants to be determined by the electromagnetic boundary conditions.
At this stage, one already sees that the ratio $C^{\rm L}_{\pm } / C_{\pm }^{\rm TM}$ is roughly equal to $q/\alpha\eta $.
This is a robust consequence of the current conservation at the boundaries and holds irrespective of the materials outside as long as they are insulating. 
Then, from Eqs.~(\ref{eq:E_metal_solved}) and (\ref{eq:Jf_result}), the ratio of the in-plane current induced by the longitudinal field to that by the transverse field is $\sim (q/\alpha ) \cdot (q/Q)$.
As concretely discussed in Sec.~\ref{sec:estimates}, typically there exists a hierarchy $\eta \ll q/\left| Q\right|  \ll \left| \alpha \right| /q \sim 1$, so that \textit{the in-plane current is induced mostly by the transverse field even when $v/c_m \ll 1$}. As for the electric field [Eq.~(\ref{eq:E_metal_solved})], the longitudinal component dominates near the interface whereas the transverse component decays slower and becomes prominent, in relative terms, into the depth of the metal, which is a different behavior from the electric current. This is because the electric current contains the diffusive component induced by the charge accumulation in addition to the drift current proportional to the electric field.
Note that these estimates can be easily confirmed for films of infinite thickness, where $C_+^{\rm L/TM/TE} = 0$, and
\begin{equation}
	\begin{aligned}
		\begin{pmatrix}
			E_x \\
			E_z \\
		\end{pmatrix}&\propto i\eta 
		\begin{pmatrix}
			i\alpha/q \\
			-1 \\
		\end{pmatrix} e^{-\alpha z}
		- \begin{pmatrix}
			iq/Q \\
			-1 \\
		\end{pmatrix} e^{-Qz} ,\\
		E_y &\propto e^{-\alpha z},
		\label{eq:inf_elec}
	\end{aligned}
\end{equation} 
\begin{equation}
	\begin{aligned}
		\begin{pmatrix}
			J_x \\
			J_z \\
		\end{pmatrix}&\propto i\eta \sigma_c \Big[
		\begin{pmatrix}
			i\alpha/q \\
			-1 \\
		\end{pmatrix} e^{-\alpha z}
		- 
		\begin{pmatrix}
			iq/Q \\
			-1 \\
		\end{pmatrix} e^{-Qz} \Big],\\
		J_y &\propto \sigma_c e^{-\alpha z}.
		\label{eq:inf_current}
	\end{aligned}
\end{equation} 
The above consideration suggests that when a conductive layer is present, regardless of the physical nature of the substrate, the components of electric currents associated with the longitudinal and transverse electric fields have comparable orders of magnitude by the constraint at the boundaries. We will clarify the implication of this observation on the quasi-electrostatic approximation in Sec.~\ref{sec:electrostatics}.

\subsection{Half-infinite piezoelectric substrate \label{sec:piezo_sols}}
In general, the analytical solutions for $z<0$ are either unavailable or intractable mainly because the piezoelectricity intrinsically lacks rotational symmetry.
Since our model ignores mechanical deformation of the conducting layer, the electromagnetic field inside the piezoelectric substrate affects the conducting layer only \textit{via} the boundary conditions, \textit{i.e.,} the amplitudes of the electric and magnetic fields at $z=-0$.
Taking advantage of this, we merely outline the framework of the numerical calculation here, avoiding the actual evaluation of the electromagnetic field and treating their interface values as input parameters in the later analysis.

The solutions can be computed by simultaneously solving Eqs.~(\ref{eq:elastic}) and (\ref{eq:TM_TE}) under the constraints (\ref{eq:Maxwell4}) and (\ref{eq:piezo_relation}) and boundary conditions described in the following~\cite{tiersten1963jasa,tseng1967jap,campbell1968ieee,ingebrigtsen1969jap}.
Eliminating $\bm{D}$ by Eq.~(\ref{eq:piezo_relation}), there exist six unknown functions, $\bm{u}$ and $\bm{E}$, governed by the six time-evolution equations (\ref{eq:elastic}) and  (\ref{eq:TM_TE}) with one constraint Eq.~(\ref{eq:Maxwell4}), implying the system should contain five dynamical degrees of freedom.
These equations are presented explicitly in Eq.~(\ref{eq:piezo_dynamical}) in Appendix~\ref{sec:app:piezo}, which shows that the highest order derivative with respect to $z$ is first for $E_z$ and second for the other five unknowns. In order to count the number of integration constants, it is more transparent to have a set of equations with a homogeneous degree of differentiation and an equal number of equations and unknowns.
One may thus algebraically eliminate $E_z $ and $\partial _z E_z$ from the seven equations (Eqs.~(\ref{eq:elastic}), (\ref{eq:TM_TE}), and Eq.~(\ref{eq:Maxwell4})) and obtain five equations of second order in $\partial _z$ for five dependent variables ($u_x ,u_y ,u_z ,E_x$, and $E_y$), solving which will yield ten integration constants. The number is reduced by half by demanding either exponential decay towards $z=-\infty $ or energy transport in the negative $z$ direction~\cite{IngebrigtsenTonning1969}.
The stress-free boundary condition \textit{i.e.}, the vanishment of the stress normal to the substrate surface, should be able to eliminate three of the five remaining constants, leaving two to be determined, which may be chosen to be the values of the $x$ and $y$ components of the electric field at the substrate surface $E_{x0} \equiv E_x \left( z\rightarrow -0 \right) $ and $E_{y0} \equiv E_y \left( z\rightarrow -0 \right)$. The rational behind this choice will become clear in the following section in which the remaining electromagnetic boundary conditions are considered.

\section{Solutions for the full film stack \label{sec:full_film}}

Now we are in a position to glue the solutions obtained above together by the electromagnetic boundary conditions.
To guide the calculations, we first review the structure of the present boundary value problem.
Per interface, there are six boundary conditions that follow from Maxwell's equations (continuity of tangential $\bm{E}, \bm{H}$ and normal $\bm{D},\bm{B}$), only four of which are independent.
This can be understood as follows.
If all the field components are independent of $y$, the $z$-components of Eqs.~(\ref{eq:Maxwell1}) and (\ref{eq:Maxwell3}),  
\begin{equation}
	\partial _t B_z = -\partial _x E_y , \quad \partial _t D_z =  \partial _x H_y - J_z ,  \label{eq:equivalence}
\end{equation}
together with the plane-wave ansatz $\partial _x \rightarrow iq$, $\partial _t \rightarrow -iqv$ and the metal-insulator boundary condition $J_z =0$ lead to  
\begin{equation}
	E_y = vB_z, \quad H_y = - vD_z 
	\label{eq:Ey=vBz}
\end{equation}
at the boundaries. Thus, the continuity of $B_z ,D_z$ is equivalent to that of $E_y , H_y$, leaving four independent conditions for each boundary ($z=0$ and $z=d$). 
In the present system, the eight conditions are imposed on the eight integration constants; $C^{\rm TM}$ and $C^{\rm TE}$ in vacuum, $C_{\pm }^{\rm TM}$ and $C_{\pm }^{\rm TE}$ in the conducting layer, and $E_{x0}$ and $E_{y0}$ in the piezoelectric substrate.
As $C_{\pm }^{\rm L}$ have been represented by $C_{\pm }^{\rm TM}$ 
[Eq.~(\ref{eq:current_solved})], they are not included here.
The resulting linear system of eight equations for eight variables is homogeneous, hence the determinant of the coefficient matrix must vanish for non-trivial solutions to exist. The physical meaning is that the SAW velocity $v$ cannot be arbitrary and has to be chosen so as to nullify the determinant. 
Any solution for $v$, which may depend on $q$, indicates existence of a branch of SAW, and there can be multiple branches or none. For each solution $v$, the linear system contains only seven independent equations, leaving one undetermined constant, the overall amplitude of the SAW mode, to be fixed by an initial condition. In typical SAW problems, the amplitude should be determined such that the energy carried by the mode matches the input power to an interdigital transducer that excites the SAW~\cite{kawada2025jap}.

The computation of $v$ can be equivalently carried out by resolving the boundary conditions one by one, which turns out to be more illuminating. Let us consider the boundary at $z=d$ between the vacuum and conductor. Two of the four boundary conditions can be used to eliminate $C^{\rm TE/TM}$ and the remaining two impose conditions among $C_{\pm }^{\rm TE}$ and $C_{\pm }^{\rm TM}$. As a result, one obtains the following two equations on $C_{\pm }^{\rm TE}, C_{\pm }^{\rm TM}$: 
\begin{align}
	\frac{\gamma }{q} &= - \frac{1}{\alpha } \frac{C_+^{\rm TE}+ C_-^{\rm TE}e^{-\alpha d}}{C_+^{\rm TE}-C_-^{\rm TE}e^{-\alpha d}} , \label{eq:TE_eq} \\
	-\frac{\varepsilon _r}{\gamma }  &= \frac{\alpha}{Q} \frac{Q \left( C_+^{\rm TM} + C_-^{\rm TM}e^{-\alpha d} \right) + iq \left( C_+^{\rm L} - C_-^{\rm L}e^{-Qd} \right)}{q\left( C_+^{\rm TM} - C_-^{\rm TM}e^{-\alpha d} \right) + i\alpha \left( C_+^{\rm L} + C_-^{\rm L}e^{-Qd} \right)}, \label{eq:TM_eq}
\end{align} 
where $C_{\pm}^{\rm L}$ can be expressed with $C_{\pm}^{\rm TM}$ \textit{via} Eq.~(\ref{eq:current_solved}). We remark that these equations are equivalent to the continuity of surface impedances $Z^{\rm TE} =E_y /H_x =  vB_z  / H_x $ and $Z^{\rm TM} = E_x /H_y =  -E_x  / vD_z $ across the boundary (see Appendix~\ref{sec:app:imp} for the detail).
Equations~(\ref{eq:TE_eq}) and (\ref{eq:TM_eq}), together with Eq.~(\ref{eq:current_solved}), offer four conditions among six constants $C_{\pm}^{\rm TE}$, $C_{\pm}^{\rm TM}$, and $C_{\pm}^{\rm L}$, which we choose to represent through two auxiliary constants, $C_0^{\rm TE}$ and $C_0^{\rm TM}$,  as 
\begin{eqnarray}
	C_{\pm }^{\rm TE} &=& \frac{C_0^{\rm TE} }{2} \left( \frac{\alpha \gamma }{q} \mp 1 \right) e^{\alpha \left( d\mp d\right) /2} ,\label{eq:metal_TE} \\
	C_{\pm }^{\rm TM} &=& \frac{C_0^{\rm TM}\alpha \eta }{2q} \Bigg\{ \left( \varepsilon _r \frac{1-i\eta }{\gamma } \pm \frac{i\alpha \eta }{q} \right) \sinh Qd \nonumber \\
	&& + \frac{q}{Q} \left( \cosh Qd - e^{\pm \alpha d} \right) \Bigg\} e^{\alpha \left( d\mp d\right) /2} ,\label{eq:metal_TM} \\
	C_{\pm }^{\rm L} &=& \pm \frac{C_0^{\rm TM} }{2}\Bigg\{ \frac{i\alpha \eta }{q}\left( e^{\pm Qd}-\cosh \alpha d\right) \nonumber \\
	&& + \left( \varepsilon _r \frac{1-i\eta }{\gamma }\mp \frac{q}{Q} \right) \sinh \alpha d \Bigg\} e^{Q\left( d\mp d\right) /2} .\label{eq:metal_L}
\end{eqnarray}

Next we move on to the boundary at $z=0$. 
Looking at Eqs.~(\ref{eq:E_metal_solved}) and (\ref{eq:metal_TE}) - (\ref{eq:metal_L}), the continuity of $E_x,E_y$ at $z=0$ is sufficient to determine $C_0^{\rm TE}$ and $C_0^{\rm TM}$ in terms of $E_{x0}$ and $E_{y0}$,
\begin{equation}
	\begin{aligned}
		\frac{E_{x0}}{C_0^{\rm TM}}=& i  \left( \frac{q^2}{Q^2} - \frac{\alpha ^2 \eta ^2}{q^2} \right) \sinh Qd \sinh \alpha d \\
		& + \frac{2\alpha \eta}{Q} \left( \cosh Qd \cosh \alpha d - 1\right) \\
		& +\varepsilon _r \frac{1-i\eta }{\gamma } \Bigg( \frac{\alpha \eta }{q} \sinh Qd \cosh \alpha d \\
		& +\frac{iq }{Q}\cosh Qd \sinh \alpha d \Bigg), \\
		\frac{E_{y0}}{C_0^{\rm TE}}=& \sinh{\alpha d} + \frac{\alpha \gamma}{q} \cosh{\alpha d}.
	\end{aligned} \label{eq:result_interface}
\end{equation}
In principle, the continuity of $H_x$ and $H_y$ at $z=0$ should give two more independent linear relations between $C_0^{\rm TM}, C_0^{\rm TE}, E_{x0}$ and $E_{y0}$, allowing one to find the value of $v$ for which there is a nontrivial solution. Gaining any analytical insight in $H_x ,H_y$ inside the piezoelectric substrate in terms of $E_{x0},E_{y0}$ seems a hopelessly complicated task, however, due to the essential anisotropy of the piezoelectric tensor. Here we forego the elasticity-related part by appealing to the empirical fact that the Rayleigh-type SAW mode exists in any material setup and is always one of the most easily excited. Thus we assume that $v$ will take a value representative of Rayleigh-type SAWs and regard the physical quantities of interest as functions of $E_{x0}$ and $E_{y0}$. The values of the electric field components at the interface can be estimated by an independent means and used to evaluate the electric current and charge distribution within a conducting layer, as explained in Sec.~\ref{sec:estimates}.

\section{Electrostatic limit}\label{sec:electrostatics}

In Sec.~\ref{sec:3b}, after Eq.~(\ref{eq:B_result}), we noted that our results reduce to the ones obtained under the quasi-electrostatic approximation if $v/c_m$ and $1/\left( q\delta _m \right)$ are both sufficiently small. On the other hand, assuming $v/c_m \ll1 $ and $1/\left( q\delta _m \right) \lesssim 1 $ in Eq.~(\ref{eq:current_solved}), we found that the electric current driven by a transverse component of electric field remains unscreened and thereby dominates over its longitudinal counterpart, seemingly suggesting inapplicability of the quasi-electrostatic approximation that is supposed to discard transverse fields. We resolve this apparent contradiction in two steps. In Sec.~\ref{sec:LTH}, we first clarify an ambiguity in the definition of transverse fields. This highlights the role of a component of the quasi-electrostatic potential, the so-called harmonic potential that satisfies $\nabla ^2 \phi =0$ and behaves like the transverse field in the exact treatment. In Sec.~\ref{sec:conditions}, we rewrite the governing equations in a conductor to demonstrate that the transverse fields become harmonic if $v/c_m \ll1 $ and $1/\left( q\delta _m \right) \ll 1$. Thereby we establish the harmonic potential as a limit of the transverse field, giving a clear physical interpretation to the unscreened electric current under the quasi-electrostatic approximation. It also shows the presence of a regime $1/\left( q\delta _m \right) \sim 1$ where the approximation fails despite $v/c_m \ll 1$. 

\subsection{Longitudinal, transverse, and harmonic fields}\label{sec:LTH}

As noted after Eq.~(\ref{eq:inf_current}), retaining the transverse fields appears essential for satisfying charge conservation at the boundaries of a conductor. This may appear at odds with the perception that the electrostatic potential should describe longitudinal fields. If this were the case, the quasi-electrostatic approximation would never work for time-dependent problems involving conductors. To resolve the puzzle, one should carefully distinguish two possible definitions of the longitudinal and transverse components. Namely, one may call $\bm{E}^\mathrm{L} $ ($\bm{E}^\mathrm{T} $) longitudinal (transverse) if \\
\indent (a) $\nabla \times \bm{E}^\mathrm{L} = \bm{0} $ ($\nabla \cdot \bm{E}^\mathrm{T} =0$), \\
or if\\
\indent (b) $\nabla \cdot \bm{E}^\mathrm{L} \neq 0 $ ($\nabla \times \bm{E}^\mathrm{T} \neq \bm{0}$).\\
So far, we have used (a) to classify longitudinal and transverse fields, which conforms with the terminology often adopted. In this way, the electric field in the quasi-electrostatic approximation is always longitudinal. However, (b) is better adapted in the screening argument as it is the condition for generating the screening charges and currents through the Gauss' and Amp\`ere-Maxwell laws, respectively. Usually, the difference between the two definitions does not matter, but there is a case that it does:  $\bm{E}=-\nabla \phi ^{\rm H}$ where the harmonic potential $\phi ^{\rm H}$ satisfies $\nabla ^2 \phi ^{\rm H}=0$. Such an $\bm{E}$ is classified as simultaneously longitudinal and transverse according to (a), but as neither of the two according to (b). When solving the Poisson's equation $\nabla ^2 \phi = -\rho /\varepsilon $ arising from Gauss' law, $\phi ^{\rm H}$ enters as a source-free solution to be added to satisfy prescribed boundary conditions. Therefore, the answer to the above question is anticipated to be that the quasi-electrostatic approximation does include a transverse field in the sense of (a), or equivalently the transverse field in the exact treatment reduces to a harmonic component in the sense of (b).

This can be seen directly by taking the limit $v/c_m \rightarrow 0 $ and $1/\left( q\delta _m \right) \rightarrow 0$, which brings the solutions presented in Sec.~\ref{sec:3b} to those of Ref.~\cite{ingebrigtsen1970jap}. One observes that
\begin{equation}
\begin{pmatrix}
E_x^{\rm TM} \\
E_z^{\rm TM} \\
\end{pmatrix} \rightarrow C_+^{'\rm TM} \begin{pmatrix}
	1 \\
	-i \\
	\end{pmatrix} e^{q\left( z-d\right)} + C_-^{'\rm TM} \begin{pmatrix}
	1\\
	i\\
	\end{pmatrix} e^{-qz} ,
	\label{eq:estaticlimit}
\end{equation}
which equals $ -i q^{-1} \nabla \left( C_+^{'\rm TM}e^{q\left( z-d\right)} +C_-^{'\rm TM}e^{-qz}\right)$ (note that $e^{iqx}$ is implicit throughout). Here $C_{\pm}^{'\rm TM}$ is the coefficient under the quasi-electrostatic approximation. We give a slightly more general derivation in the next subsection.

This confusion in semantics can result in a false preconception that electric fields in the electrostatic regime are longitudinal and should therefore be screened efficiently by charges.
The first and second parts of the reasoning are correct only under the definitions (a) and (b), respectively. Abandoning the quasi-electrostatic approximation, the present work demonstrates that the unscreened harmonic component, established in a separate publication~\cite{kawada2024exparxiv} as the origin of the spin current in Ref.~\cite{kawada2021sciadv}, should in fact be considered as a transverse field screened by magnetically induced currents in the limit of infinite skin depth $1/\left( q\delta _m \right) \rightarrow 0$.

\subsection{Conditions for quasi-electrostatic approximation}\label{sec:conditions}

The previous subsection suggests that for the quasi-electrostatic approximation to hold, the difference between the harmonic component in it and the transverse field without the approximation should be small. Below, we obtain a quantitative condition for this. We derive decoupled bulk equations for longitudinal and transverse components, identify the boundary condition that couples them, and compare them to the equations for the quasi-electrostatic approximation. 

We begin with the Helmholtz decomposition of the electric field $\bm{E} = -\nabla \phi + \bm{E}^{\rm T} , \nabla \cdot \bm{E}^{\rm T}=0$. The quasi-electrostatic approximation is implemented by setting $\bm{E}^{\rm T}=\bm{0}$. While such $\phi $ and $\bm{E}^{\rm T}$ always exist at least locally, the decomposition is not necessarily unique since $\phi + \phi ^{\rm H}, \bm{E}^{\rm T}+ \nabla \phi ^{\rm H}$ with $\nabla ^2 \phi ^{\rm H}=0$ would give another. This ambiguity has no physical consequence as long as both the potential and transverse field are retained and consistently shifted. 
In our exact treatment, the solution Eq.~(\ref{eq:E_metal_solved}) does not contain nonzero $\phi ^{\rm H}$ so that we do not have to distinguish definitions (a) and (b) of the longitudinal and traverse fields.

Starting from Eqs.~(\ref{eq:Maxwell1}) - (\ref{eq:Maxwell4}) with $\bm{B}=\mu _0 \bm{H}$, we first exclude the time-independent component of $\bm{B} $ from the discussion, which can be separately characterized by the \textit{magnetostatics}. Then, Eq.~(\ref{eq:Maxwell2}) can be disregarded and Eq.~(\ref{eq:Maxwell1}) is used to eliminate $\bm{B},\bm{H}$. Differentiating Eq.~(\ref{eq:Maxwell3}) in time, one obtains
\begin{equation}
\nabla ^2 \bm{E}^{\rm T} =\varepsilon \mu _0 \partial _t^2 \left( \bm{E}^{\rm T}-\nabla \phi \right) + \mu _0 \partial _t \bm{J} .
\end{equation}
Specializing in the isotropic conductor, using Eqs.~(\ref{eq:Maxwell4}) and (\ref{eq:isotropic_conductor}) yields
\begin{eqnarray}
&& \left( \frac{\partial _t^2}{c_m^2} +\mu _0\sigma _c \partial _t -\nabla ^2 \right) \bm{E}^{\rm T} \nonumber \\
&=& \left\{ \frac{\partial _t^2}{c_m^2} + \mu _0 \left( \sigma _c -D_e \varepsilon \nabla ^2 \right) \partial _t \right\} \nabla \phi \equiv \bm{w} . \label{eq:wave} 
\end{eqnarray}
One expects $\bm{E}^{\rm T}$ and $\phi $ to decouple, for which $\bm{w}$ should vanish. By taking curl and divergence of Eq.~(\ref{eq:wave}), one deduces $\bm{w} = \nabla \chi $ where $\nabla ^2 \chi =0$. Even if $\chi \neq 0$, one may always eliminate it by adjusting the Helmholtz decomposition with a $\phi ^{\rm H}$ such that 
\begin{equation}
    -\left( \frac{\partial _t^2}{c_m^2}+\mu _0 \sigma _c \partial _t  \right) \nabla \phi ^{\rm H} = -\nabla \chi ,
\end{equation}
which can always be solved without affecting physical quantities. Therefore without loss of generality, Eq.~(\ref{eq:wave}) reduces to
\begin{eqnarray}
\left( \frac{\partial _t^2}{c_m^2} +\mu _0 \sigma _c \partial _t -\nabla ^2 \right) \bm{E}^{\rm T} &=& 0 , \label{eq:transverse_wave} \\
\left\{ \frac{\partial _t }{c_m^2} + \mu _0 \left( \sigma _c -D_e \varepsilon \nabla ^2 \right) \right\} \nabla \phi &=& 0 ,\label{eq:longitudinal_wave}
\end{eqnarray}
where we excluded static (or unbounded in time) contributions in deriving Eq.~(\ref{eq:longitudinal_wave}).

In a conducting film, the evanescent wave associated with SAW has both longitudinal and transverse components so that there must be a mechanism that couples them. It primarily arises from the boundary conditions. Specifically, the insulating condition $\bm{J}\cdot \bm{n} =0$ at the conductor surface with $\bm{n}$ the surface normal gives \textit{via} the Gauss' law and drift-diffusion relation
\begin{equation}
\bm{n}\cdot \bm{E}^{\rm T} = \left( 1-\frac{\varepsilon D_e \nabla ^2}{\sigma _c} \right) \bm{n}\cdot \nabla \phi = -\frac{\partial _t}{\mu _0 c_m^2 \sigma _c}  \bm{n}\cdot \nabla \phi . \label{eq:boundary_transverse} 
\end{equation}
The second equality follows from Eq.~(\ref{eq:longitudinal_wave}). If one were to assume a purely longitudinal field in the layer geometry, one would have $\partial _z \phi =0$ at both surfaces. 
Such solutions of Eq.~(\ref{eq:longitudinal_wave}) would exist only for some special values of $v$, determined independently of the acoustic wave properties. Therefore, at least on one surface, it is expected that $\bm{E}^{\rm T}$ is of the same order as $\eta \partial _z \phi $. This is a restatement of Eq.~(\ref{eq:current_solved}).

In the quasi-electrostatic approximation, on the other hand, one sets $\bm{E}^{\rm T} =\bm{0}$ so that there is no freedom to shift out $\phi ^{\rm H}$ from $\phi $ and the harmonic component may have a physical significance. Equation~(\ref{eq:Maxwell3}) cannot be satisfied, meaning Eq.~(\ref{eq:wave}) is unavailable, and one relies instead on
\begin{equation}
    \left\{ \frac{\partial _t}{c_m^2}+\mu _0 \left( \sigma _c -D_e \varepsilon \nabla ^2 \right) \right\} \rho = 0, \label{eq:continuity}
\end{equation}
derived from the charge conservation, drift-diffusion relation, and Gauss' law $-\varepsilon _r \varepsilon _0 \nabla ^2 \phi = \rho$. Any harmonic potential with arbitrary time dependence is its trivial solution and can be added to any solution. For a solution $\phi ^{\rm L}$ of Eq.~(\ref{eq:continuity}) with $\nabla ^2 \phi ^{\rm L} \neq 0$, which therefore also solves Eq.~(\ref{eq:longitudinal_wave}), the zero current boundary condition yields
\begin{equation}
-\bm{n}\cdot \nabla \phi ^{\rm H} = \left( 1-\frac{\varepsilon D_e \nabla ^2}{\sigma _c} \right) \bm{n}\cdot \nabla \phi ^{\rm L} . \label{eq:boundary_harmonic}
\end{equation}
This appears to be the only constraint on the time dependence of $\phi ^{\rm H}$. Comparing Eqs.~(\ref{eq:boundary_transverse}) and (\ref{eq:boundary_harmonic}), one concludes that the harmonic component in the quasi-electrostatic approximation plays the role of $\bm{E}^{\rm T}$ in the exact treatment.

Looking at Eq.~(\ref{eq:transverse_wave}), one observes that if the time-derivatives can be ignored, which in the case of plane waves reduces to $v^2 /c_m^2 \ll1 $ and $ 1/\left( q\delta _m \right) ^2 \ll 1$, the transverse field becomes harmonic $\nabla ^2 \bm{E}^{\rm T} \rightarrow \bm{0}$. Since $\nabla \cdot \bm{E}^{\rm T}=0$ by definition, it implies $\nabla \times \bm{E}^{\rm T}=\bm{0}$, and hence $\bm{E}^{\rm T} =-\nabla \phi ^{\rm H}$ with some harmonic potential $\phi ^{\rm H}$. This formalizes the discussion at the end of the previous subsection and establishes the validity conditions for ignoring $\bm{E}^{\rm T}$ as well as its relation to $\phi ^{\rm H}$.

\section{Quantitative estimation}\label{sec:estimates}

In this section, we calculate the profile of the electric current, charge density, and magnetic field in the conducting layer. 
We model a system in which SAW-induced spin current was observed experimentally~\cite{kawada2021sciadv}. 
Here material parameters that match the experimental conditions are used. We discuss the consequence of the results obtained here on the SAW-induced spin current.

\begin{figure}[tb]
	\begin{minipage}{1.0\hsize}
		\centering
		\includegraphics[scale=0.16]{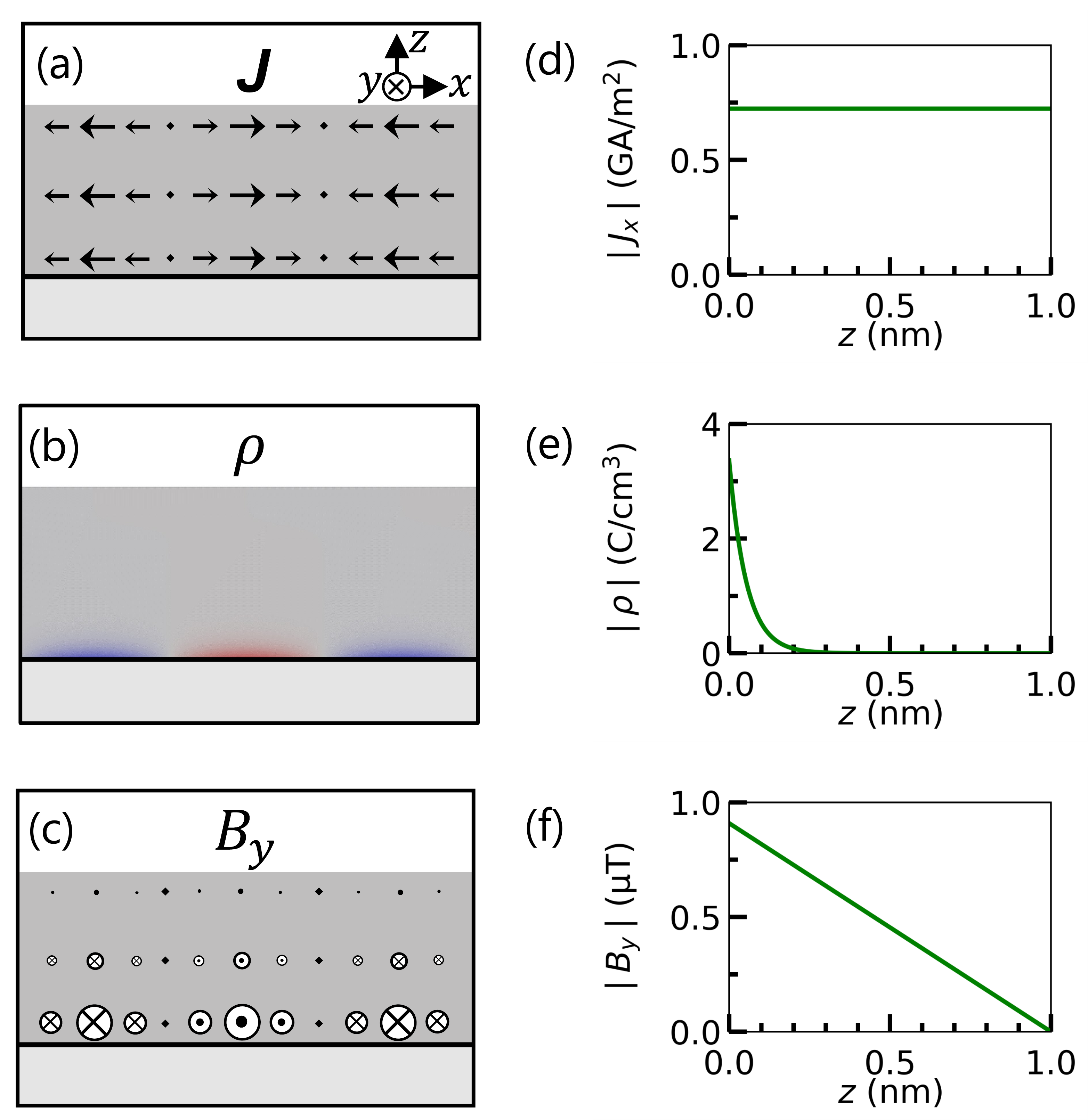}
	\end{minipage}
	\caption{
		(a-c) Schematic illustration of $(J_x,J_z)$	(a), $\rho$ (b), and $B_y$ (c) within the metallic thin film. The light and dense gray regions represent the piezoelectric substrate and the metallic thin film, respectively. The coordinate system is shown in (a), and the boundary of the two regions is situated at $z=0$. $J_z$ is almost invisible in (a) since it is several orders of magnitude smaller than $J_x$. In (b), positive (negative) induced charge density is expressed with red (blue).
		(d-f) Profile of $\abs{J_{x}}$ (d), $\abs{\rho}$ (e), and $\abs{B_y}$ (f) along $z$-axis. 
		$(J_x,J_z)$, $\rho$, and $B_y$ are estimated from Eqs.~(\ref{eq:Jf_result}), (\ref{eq:rhof_result}), and (\ref{eq:B_result}) with $\sigma_c=10^6\ /\qty(\mathrm{\Omega\cdot m})$, $\varepsilon_r=1$, $d=1$ nm, $v=3871$ m/s, $q = 2\pi / 10\ \upmu$m$^{-1}$, and $\abs{E_{x0}}=10$ V/cm. 
		Note that $J_y$ and $(B_x,B_z)$ will also be induced in the TE mode.
		\label{fig:calc_Fig4_rev}
	}
\end{figure}

In accordance with the experimental conditions, we consider a SAW propagating on a Y$+128^\circ$-cut LiNbO$_3$ substrate along the crystal $X$ axis, with the surface coated by a metallic thin film. The conductivity, relative permittivity, and thickness of the film are specified as $\sigma_c = 10^6\ (\mathrm{\Omega \cdot m})^{-1}$, $\varepsilon_r = 1$, and $d = 1$ nm, respectively. The diffusion coefficient is chosen to be $D_e = 3\ \mathrm{cm^2/s}$, a representative value for metals~\cite{maekawa2017book,niimi2013prl}. The SAW wavenumber along the $x$ direction is taken as $q = 2\pi/10\ \upmu\mathrm{m}^{-1}$. It is reasonable to assume that the SAW velocity is on the order of $10^3$ m/s, yielding a skin depth of several tens of micrometers. This condition may justify the application of the quasi-electrostatic approximation. Then, using these parameters, we numerically evaluate the SAW propagating through the full film stack within the quasi-electrostatic approximation~\cite{ingebrigtsen1970jap,kawada2024exparxiv,kawada2025jap}, obtaining a SAW velocity of $v = 3871$ m/s. From the same parameter set, we derive $\lambda_\mathrm{TF} = 0.054$ nm, $q/|Q| \sim 10^{-5}$, $\eta \sim 10^{-8}$, $v/c_m = 1.3 \times 10^{-5}$, and $\delta_m = 25\ \upmu$m. These results confirm that $v/c_m \ll 1$ and $1/(q \delta_m) \ll 1$, thereby validating the quasi-electrostatic approximation. It is further noted that qualitatively similar results are obtained as long as the film thickness satisfies $d \lesssim 1/|\alpha| \sim 100$ nm.

For the parameters given above, $\left| \alpha \right| d \sim qd \sim 10^{-3}$ and $\left|Q \right|d \sim 20$ hold so that $\left|\sinh \alpha d \right|\sim 10^{-3}$, $\left|\cosh \alpha d \right|\sim 1$, $\left|\sinh Qd \right|\sim\left|\cosh Qd\right|\gg 1$.
Therefore, Eq.~(\ref{eq:result_interface}) is well approximated by 
\begin{equation}
	C_{0}^{\rm TM} \sim \frac{2e^{-Qd}}{\varepsilon _r q d} \frac{E_{x0}}{\eta /qd + iq/Q}, \quad C_{0}^{\rm TE} \sim E_{y0}.
\end{equation}
Note that $\eta$ cannot be set to zero even at this leading order.
Using the same approximations, one obtains
\begin{eqnarray}
	\begin{pmatrix}
		E_x^{\rm TM} \\
		E_z^{\rm TM} \\
	\end{pmatrix}
	&\sim& \frac{ \eta }{qd}
	\frac{ E_{x0} }{\eta /qd + iq/Q} 
	\begin{pmatrix}
		1 \\
		iq(d - z) \\
	\end{pmatrix}, \\
	\begin{pmatrix}
		E_x^{\rm L} \\
		E_z^{\rm L} \\
	\end{pmatrix}
	&\sim &
	\frac{ E_{x0} e^{-Qz} }{\eta /qd + iq/Q} 
	\begin{pmatrix}
		iq/Q  \\
		-1 \\
	\end{pmatrix}.
\end{eqnarray}
As shown in Eqs.~(\ref{eq:Jf_result}) and (\ref{eq:rhof_result}), $\bm{E}^{\rm L}$ is effectively screened by the small factor $\eta$ while $\bm{E}^{\rm TM}$ remains sizable in the film.
Therefore, the electric current inside conducting layer is completely dominated by the transverse contribution whereas the charge density is induced by $\bm{E}^{\rm L}$. We note that 
$\left| E_x^{\rm TM} \right| \gg \left| E_z^{\rm TM} \right| $ is consistent with $\nabla \cdot \bm{E}^{\rm TM} =0$ as the smallness of $E_z^{\rm TM}$ is a result of cancellation between the $C_+^{\rm TM}$ and $C_-^{\rm TM}$ terms; such cancellation is absent in the $z$ derivative $\partial _z E^{\rm TM}_z$ because of the sign change of the latter term. 

The uniformity of the electric current, the localization of the charge density, and the gradient of the magnetic field within the conducting layer are directly demonstrated by evaluating Eqs.~(\ref{eq:Jf_result}-\ref{eq:B_result}) using Eq.~(\ref{eq:E_metal_solved}). 
The results are schematically illustrated in Fig.~\ref{fig:calc_Fig4_rev}(a-c).
Due to the small factor of $\eta$ originating from the electrostatic screening, the longitudinal components of $\bm{J}$ associated with $\bm{E}^{\rm L}$ localized at the film surfaces become negligibly small.
In addition, the cancellation between the $C_{+}^{\rm TM}$ and $C_{-}^{\rm TM}$ terms results in a significant decrease of $J_z$ overall.
The charge density $\rho$ is confined near the interface between the substrate and film to screen out $E^{\rm L}_z$ emanating from the substrate.
The TM component of the magnetic flux density $B_y$ becomes vanishingly small near the film/vacuum interface owing to $B_y \propto E_z^{\rm{TM}} \sim J_z$ (see Eq.~(\ref{eq:B_result})) and the insulating condition at $z=d$. Note that contribution of $\bm{E}^{\rm L}$ to $J_z$ survives at $z=0$, leading to non-vanishing $B_y$ there.
To illustrate the distribution along $z$ more quantitatively, the absolute values of $J_x$, $\rho$, and $B_y$ are plotted against $z$ in Fig.~\ref{fig:calc_Fig4_rev}(d-f). Here we set $\abs{E_{x0}}=10$ V/cm, which corresponds to SAW strain $|\epsilon_{xx}|=2\times 10^{-4}$ and is thought to be a practical value for the present setup judging from the calculation result to determine $v$~\cite{ingebrigtsen1970jap,kawada2024exparxiv,kawada2025jap}.
We see that $J_x$ is almost uniform along $z$ in the whole film, $\rho$ is concentrated within the Thomas-Fermi length from the substrate surface, and $B_y$ decreases linearly towards the film surface.
Since the estimated $B_y$ is comparable to the Barnett field due to the SAW-induced rotational motion of surface atoms~\cite{kurimune2020prl,tateno2021prb}, the magnetic field associated with the SAW can significantly influence the magnetization dynamics in ferromagnetic thin films with weak magnetoelastic coupling such as permalloy. Although $B_y$ is about one order of magnitude smaller than the typical magneto-elastic effective field~\cite{dreher2012prb}, the directions of the two are different and thus angle dependence of the magnetoacoustic resonance spectrum can be modified. In addition, the gradient of the magnetic field can generate the diffusive spin current along the film thickness direction, which has the same geometrical configuration as the spin current generation \textit{via} spin-vorticity coupling~\cite{matsuo2013prb,kobayashi2017prl}. 

Recent experimental work has demonstrated that SAWs induce an ac spin current in metallic thin films with significant spin-orbit interaction, an effect referred to as the acoustic spin Hall effect~\cite{kawada2021sciadv}. The observed spin current flows parallel to the surface normal with its polarization orthogonal to both the flow and SAW propagation directions, and is distributed uniformly across the film thickness direction. Under the assumption that electric fields are efficiently screened by conduction electrons and decay rapidly in metals, the latter feature suggests that the spin current should have a mechanical origin rather than electromagentic one. From this work, however, we conclude that the SAW-induced electric field generates an electric current uniformly in the thickness direction in metallic thin films. This allows the uniform spin current to flow in strong spin-orbit metals \textit{via} the spin Hall effect, which can reasonably explain the microscopic origin of the acoustic spin Hall effect.

\section{Conclusion}
In summary, we have studied the electromagnetic response of a metallic thin film to the electric field associated with the piezoelectrically excited surface acoustic wave (SAW) without employing the quasi-electrostatic approximation.
The electromagnetic field accompanying the SAW contains a component that behaves as an evanescent field and carries a transverse field, which is screened not by electric charge but by the skin effect inside conductors. 
The transverse electric field induces a uniform electric current inside the conducting layer, provided that it is thinner than the skin depth.
The transverse electric field generates an evanescent magnetic field, whose magnitude can be comparable to effective field induced by the magneto-elastic coupling and/or the spin-vorticity coupling.
The electromagnetic evanescent field that accompanies SAW can therefore induce carrier transport in metallic thin films deposited on piezoelectric substrates, a playground for studies on acoustoelectronics and spintronics.

\section{Acknowledgements}
We acknowledge fruitful discussions with K. Usami. This work was partly supported by JSPS KAKENHI (Grant Nos. 20J21915, 20J20952, 21K13886, 23KJ1419, 23KJ1159, 23H05463, and 24K00576), JST PRESTO Grant No. JPMHPR20LB, Japan, and JSPS Bilateral Program Number JPJSBP120245708.

\section{Appendix}
\subsection{\label{sec:app:piezo} Explicit form for half-infinite piezoelectric substrate} 

Here we explicitly describe equations for SAWs in a piezoelectric substrate. To emphasize the algebraic structure, let us introduce three-by-three matrices
\begin{equation}
	\begin{gathered}
		G_{ij} = c_{i1j1} ,\quad S_{ij} = c_{i3j3} , \quad T_{ij} =  c_{i3j1} ,\\
		X_{ij} = e_{ij1} , \quad Y_{ij} = e_{ij3} .
	\end{gathered}
\end{equation}
Equations~(\ref{eq:elastic}) and (\ref{eq:TM_TE}) under the constraints Eqs.~(\ref{eq:Maxwell4}) and (\ref{eq:piezo_relation}) can be written as
\begin{widetext}
	\begin{align}
	\begin{pmatrix}
		\partial _z^2 \hat{S} + iq\partial _z \left( \hat{T} +\hat{T}^T \right)  -q^2 \hat{G} -\rho _p q^2 v^2  \hat{I} & - iq\hat{X}^T -\partial _z \hat{Y}^T \\
		\frac{iq\hat{X} + \partial _z \hat{Y}}{\varepsilon _0} & \partial _z^2 \hat{I}_3 -iq \partial _z \hat{I}_2 + q^2 \left( \frac{\hat{\varepsilon }}{\varepsilon _0}\frac{v^2}{c^2} -\hat{I}_1 \right)  \\
	\end{pmatrix} \begin{pmatrix}
		\bm{u} \\
		\bm{E} \\
	\end{pmatrix} &= 0 , \label{eq:piezo_dynamical}  \\
	\left\{ \partial _z^2 Y_{3j} +iq\partial _z \left( X_{3j} + Y_{1j} \right) -q^2 X_{1j} \right\} u_j + \left( \partial _z \varepsilon _{3j} +iq\varepsilon _{1j} \right) E_j &= 0 ,\label{eq:piezo_constraint}
	\end{align}
\end{widetext}
where $\hat{I}$ is the three-by-three unit matrix and
\begin{equation}
	\hat{I}_1 = \begin{pmatrix}
		0 &0 & 0 \\
		0 & 1 & 0 \\
		0 & 0 & 1 \\
	\end{pmatrix} , \quad \hat{I}_2 = \begin{pmatrix}
		1 & 0 & 0 \\
		0 & 0 & 0 \\
		0 & 0 & 1\\
	\end{pmatrix}, \quad \hat{I}_3 = \begin{pmatrix}
		1 & 0 & 0 \\
		0 & 1 & 0 \\
		0 & 0 & 0 \\
	\end{pmatrix} .
\end{equation}
Since Eq.~(\ref{eq:piezo_dynamical}) does not contain second order derivative on $E_z$, the mathematical structure of the problem is harder to see than the other cases. 
One formal approach is to eliminate $E_z$ and $\partial _z E_z$ by using the last of Eq.~(\ref{eq:piezo_dynamical}) and Eq.~(\ref{eq:piezo_constraint}), which should leave an equation for $\bm{u} ,E_x ,E_y$ of second order in $\partial _z$.
Although this procedure cannot be used to eliminate $E_z $ when $\varepsilon _{33} = \varepsilon _{13} =0$, it is applicable for the Rayleigh-type SAW, the system of our interest. Then, one expects to obtain five independent solutions of the form $\propto e^{kz} , (k>0)$, and the stress-free boundary condition
\begin{equation}
	\left( \partial _z \hat{S} + iq \hat{T} \right) \bm{u} -\hat{Y}^T \bm{E} = 0 
\end{equation}
should be able to eliminate three of the five arbitrary constants, leaving two to be determined through the electromagnetic boundary conditions. 

\subsection{\label{sec:app:imp} Surface impedance at substrate/film interface}
In the main text, we expressed the electric field inside the conducting layer in terms of the electric field at the substrate/film interface.
As we mentioned in the last paragraph of Sec.~\ref{sec:full_film}, obtaining $v$ and the relative amplitude between $E_{x0}$ and $E_{y0}$ requires the continuity of $H_x$ and $D_z$ at $z=0$. The conditions are equivalent to equating the surface impedance of both sides, as described by Refs.~\cite{ingebrigtsen1969jap,ingebrigtsen1970jap}. The surface impedance is defined in each region independently. We first derive the analytical formula of the surface impedances in the conducting layer, which read
\begin{equation}
	\begin{aligned}
		Z^{\rm TE} = &- i\sqrt{\frac{\mu _0}{\epsilon _0}} \frac{v}{c} \frac{q}{\alpha } \frac{C_+^{\rm TE}e^{-\alpha d} +C_-^{\rm TE}}{C_+^{\rm TE}e^{-\alpha d} - C_-^{\rm TE}} , \\
		Z^{\rm TM} =& - i \sqrt{\frac{\mu _0}{\epsilon _0}}\frac{c}{\varepsilon _r v} \\
		& \times  \frac{\alpha \left( C_+^{\rm TM} e^{-\alpha d} - C_-^{\rm TM} \right) -iq \left( C_+^{\rm L} e^{-Qd} +C_-^{\rm L} \right)}{q\left( C_+^{\rm TM}e^{-\alpha d} +C_-^{\rm TM} \right) -i Q \left( C_+^{\rm L}e^{-Qd} -C_-^{\rm L}\right)} .
	\end{aligned}
\end{equation}
From Eqs.~(\ref{eq:metal_TE}) and (\ref{eq:metal_TM}), one obtains the surface impedance for the conducting layer as
\begin{equation}
	Z^{\rm TE} = i\sqrt{\frac{\mu _0}{\epsilon _0}} \frac{v}{c} \frac{q}{\alpha } \frac{q \tanh \alpha d + \alpha \gamma  }{q + \alpha \gamma \tanh \alpha d }, \label{eq:ZTE}
\end{equation}
\begin{widetext}
	\begin{align}
		Z^{\rm TM} =& i \sqrt{\frac{\mu _0}{\varepsilon _0}}\frac{c}{v} \frac{1}{\varepsilon _r \left( 1-i\eta \right)} \frac{\alpha }{q} \Bigg[ \left( \frac{q}{Q}\cosh Qd\sinh \alpha d -\frac{i\alpha \eta }{q} \sinh Qd \cosh \alpha d\right) \frac{\varepsilon _r \left( 1-i\eta \right)}{\gamma }  + \left( \frac{q^2}{Q^2}-\frac{\alpha ^2 \eta ^2}{q^2} \right) \sinh Qd \sinh \alpha d  \nonumber \\
		& + \frac{2i\alpha  \eta }{Q} \left( 1-\cosh Qd \cosh \alpha d\right) \Bigg] \Bigg/  \left[ \left\{ \frac{\varepsilon _r \left( 1-i\eta \right) }{\gamma } \sinh Qd +\frac{q }{Q} \cosh Qd \right\} \sinh \alpha d -\frac{i\alpha  \eta }{q} \sinh Qd \cosh \alpha d \right].\label{eq:ZTM}
	\end{align}
\end{widetext}
$Z^{\rm TM}$ agrees with the expression given in Ref.~\cite{ingebrigtsen1970jap} when setting $\alpha =q $ and $\gamma =1$.
The expression of $Z^{\rm TE}$ can be derived only when the electrostatic approximation $v/c\rightarrow 0$ is not employed.
Assuming $q\delta _m$ is neither very small nor large, $Z^{\rm TM}$ involves two small parameters; $q/Q$ and $\eta $. In any case, the dominant contributions arise from the terms proportional to $\varepsilon _r$ in both the numerator and denominator. We also remark that the limit $\sigma _c \rightarrow \infty $ implies $Q\rightarrow \infty , \eta \rightarrow 0$, which yields a well-defined limit and $Z^{\rm TM}=0$ consistent with the so-called shorted boundary condition~\cite{campbell1968ieee}.

Next we outline how the surface impedance of the adjoining piezoelectric medium can be obtained. 
The general structure of the bulk solution of the electromagnetic fields can be inferred as discussed in Sec.~\ref{sec:piezo_sols}. This solution should contain two arbitrary constants that characterize it (for instance, $E_{x0}$ and $E_{y0}$), allowing for the numerical computation of the surface impedance $Z^{\rm TE}$, $Z^{\rm TM}$ on the side $z < 0$ if the value of $v$ is determined. 

\bibliography{reference_ASHE}

\end{document}